\documentclass[amsmath,amssymb,aps,floats,twocolumn]{revtex4}

\usepackage{graphicx,subfigure,subeqnarray,fancyhdr,amsmath,epstopdf,multirow,color, mathrsfs, amssymb}
\def\r{\mathbf{r}}\def\v{\mathbf{v}}\def\t{\mathbf{t}}
\def\f{\mathbf{f}}\def\I{\mathbf{I}}\def\m{\mathbf{m}}\def\R{\mathbf{R}}
\def\d{\mathbf{d}}\def\e{\mathbf{e}}
\def\z{\mathbf{e}_z}
\def\ra{^{(1)}}\def\rb{^{(2)}}\def\rab{^{(1)\to(2)}}\def\rba{^{(2)\to(1)}}
\def\ri{^{(i)}}\def\rj{^{(j)}}\def\h{{\bf h}}
\renewcommand{\eqref}{Eq.~\originaleqref}

\def \I{\mathbf{I}}

\def \0{\mathbf{0}}
\def \R{\mathbf{R}}

\def\dd{{\rm d}}
\begin{document}

\title{Hydrodynamic interactions between nearby slender filaments}

\author{Yi Man}
\author{Lyndon Koens}

\author{Eric Lauga}
\email{e.lauga@damtp.cam.ac.uk}
\affiliation{Department of Applied Mathematics and Theoretical Physics, Centre for Mathematical Science,
  University of Cambridge, Wilberforce Road, Cambridge CB3 0WA, United Kingdom}
\date{\today}

\begin{abstract}
Cellular biology abound with  filaments interacting through  fluids, from  intracellular  microtubules, to rotating flagella and  beating cilia. While previous work has demonstrated the complexity of capturing  nonlocal hydrodynamic interactions between moving filaments, the problem  remains difficult theoretically. We show here  that when filaments are closer to each other than their relevant length scale, the integration of hydrodynamic interactions can be approximately carried out analytically. This leads to a set of simplified local equations, illustrated on a simple model of  two interacting filaments, which can be used to tackle theoretically a range of problems in biology and physics.
\end{abstract}
\maketitle

While one tends to think of  biological cells as stubby, their environment is in fact rich with filamentous structures. Inside cells, polymeric filaments  of microtubules, actin, and intermediate filaments fill the eukaryotic cytoplasm \cite{alberts} and provide it with its mechanical structure \cite{boal_book}. Outside cells, the motion of flagella and cilia allows cells to generate propulsive forces \cite{Berg73,brennen77,lauga09} and induces flows critical to  human health \cite{sleigh88,fauci06}.

In all cases, these biological filaments are immersed in a viscous fluid in which they move  at low Reynolds number, be it due to their polymerisation, to fluctuations and thermal forces, or to the action of molecular motors \cite{braybook}. At low Reynolds number, the flows  induced locally by the  motion of filaments relative to a background fluid have a slow spatial decay as  $\sim 1/r$ \cite{lighthill75,leal}. In  situations where filaments are close to each other, we thus expect nonlocal hydrodynamic interactions  to be important \cite{goldstein2016elastohydrodynamic}.

Integrating long-ranged hydrodynamic interactions between filaments has long been recognised as a challenging problem, and one where the theoretical approach has  consisted of either full numerical simulations or very simplified analysis. A variety of computational methods have been developed to tackle it including  slender-body theory \cite{johnson80,gueron93,tornberg04},  boundary elements to implement boundary integral formulations \cite{kanehl14},    the immersed boundary method \cite{lim12,yang08},   regularised flow singularities \cite{flores05} and   particle-based methods \cite{yang10,reigh12}. 

While these computational approaches allow to address complex  geometries and dynamics, the difficulty of integrating long-range hydrodynamic interactions has prevented  analytical approaches  from providing insight beyond simplified setups. The two most {common approaches in biophysics} consist in replacing the dynamics in three dimensions by a two-dimensional problem for which the analysis may be easier to carry out   \cite{Taylor51,Elfring09}, or by focusing on far-field hydrodynamic interactions and ignoring the geometrical details of  near-field hydrodynamics, a popular approach to study synchronisation of  flagella and cilia \cite{Vilfan06,guirao07,Niedermayer08,uchida10,golestanian11,friedrich12,bennett13}.

\begin{figure}[t]
\centering
\includegraphics[width=0.45\textwidth]{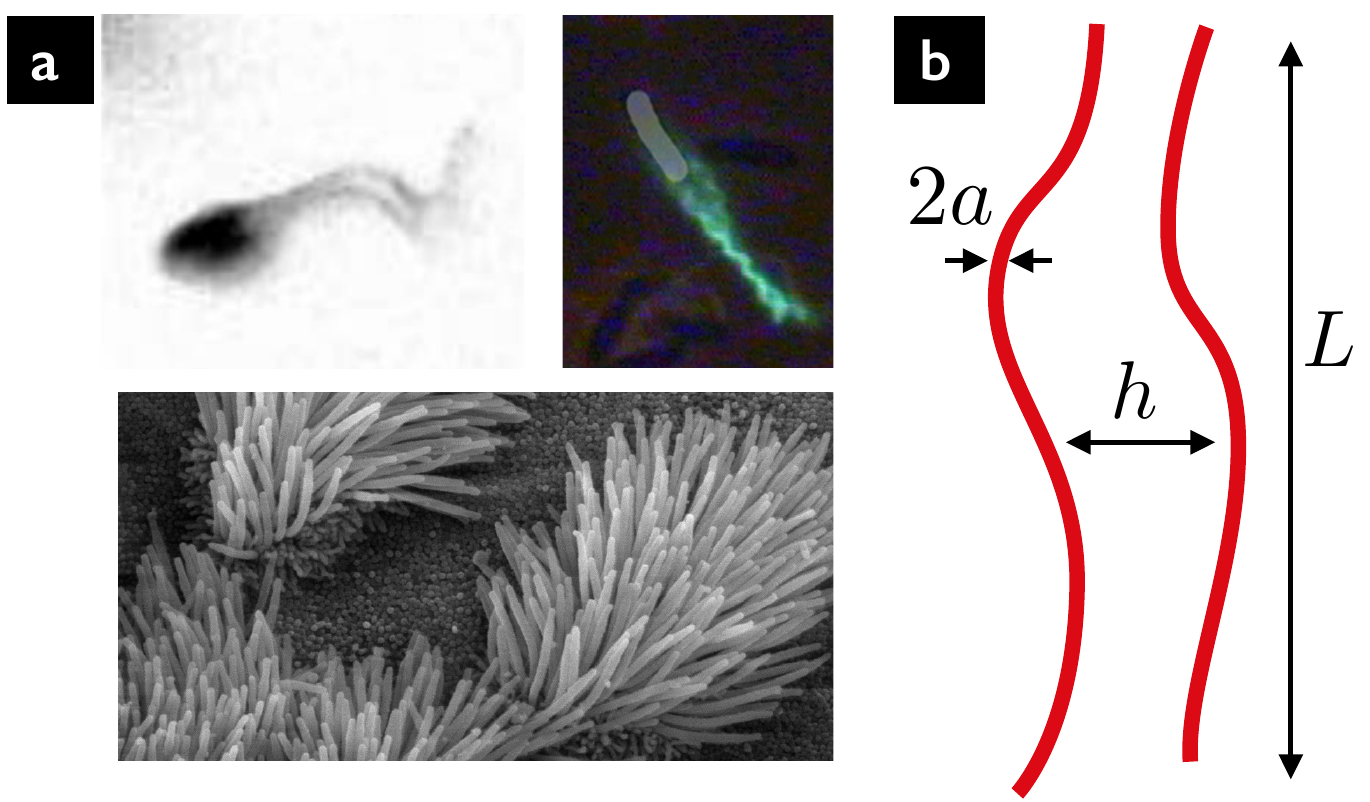}
\caption{(a) Examples where nearby filaments interact through a viscous fluid (clockwise from top left): two spermatozoa synchronising their flagella (reprinted with permission from Yang, Elgeti \& Gompper,  {\it Physical Review E}, {\bf 78}, 061903, 2008~\cite{yang2008}; copyright 2008, American Physical Society); flagellar filaments of peritrichous bacteria during swarming  (reprinted with permission from Turner, Zhang, Darnton \& Berg, {\it Journal of Bacteriology}, {\bf 192}, 3259--3267, 2010~\cite{turner10}; copyright 2010,   American Society for Microbiology);  epithelium cilia (courtesy of C.~Daghlian, Wikimedia Commons). (b) Prototypical setup: two slender filaments of length $L$ and radius $a$  at a typical distance $h$ from each other.}\label{fig1}
\end{figure}

In realistic biological situations, three-dimensional filaments are not far from each other, but in fact are often found in the opposite, near-field, limit where their separation   distance is  much smaller than their length.  This is illustrated in Fig.~\ref{fig1}a with three examples relevant to cell motility: synchronising flagella of spermatozoa; bundle of  bacterial flagellar filaments;  epithelium cilia.  In order to capture the dynamics of these  interacting filaments, new analytical tools are thus required. 

In this paper, we show that analytical progress can be achieved by taking advantage of a separation of length scales.  A generic two-filament setup  (as in Fig.~\ref{fig1}b)  is characterised by three length
 scales: the filament radius, $a$;  the separation distance between the filaments, $h$; and the  filament length, $L$. While far-field studies focus on the limit $h\gg \{L,a\}$,  many biological  situations are in the opposite near-field limit,  for example in the case of   waving cilia arrays  \cite{blake74}, for which $ \{a,h\} \ll L$, i.e.~slender filaments close to each other compared to their typical size. We show here that in the special case  where $a \ll h$, 
i.e.~for filaments thinner than any another other length scale in the problem, the hydrodynamic  interactions between the filaments can be analytically integrated out, leading to a set of simplified local equations valid in  the limit $a\ll h \ll L$. 
Our results, illustrated on a simple model of two interacting rigid filaments,  will allow to tackle theoretically a range of problems in biology and  physics.

Consider the two filaments in Fig.~\ref{fig1}b, numbered \#1 and \#2. Denote the location of the centerline to filament $i$ as $\r^{(i)}(s,t)$ where $s$ is the arclength, and let  $\t^{(i)}(s)={\partial\r^{(i)}}/{\partial s}$
 be  its unit  tangent.  {In order to compute the hydrodynamic forces on the filaments, we exploit the two assumed separations  of scales, $a\ll h \ll L$. We first note that the limit $a\ll L$  implies that the filaments are slender. Furthermore, since the  displacements of the filaments are at most on the order of their separation distance, $h$, their typical curvature, denoted $\kappa$, is at most of order $\kappa\sim h/L^2$. Since we assume the limit $h \ll L$, this means that we have always  $\kappa h\ll 1$ and $\kappa L\ll 1$, and the filaments undergo long-wavelength deformation. In that case, resistive-force theory may be used  to calculate the 
 hydrodynamic force densities on each filament \cite{gray55,cox70,lighthill75}. Denoting the force densities   $\f\ra$ and $\f\rb$, resistive-force theory states that they  are proportional to the local velocity of the filament  relative the background fluid i.e.}
\begin{subeqnarray}\label{fh}
\f\ra=-\left( \xi_\perp\I + (\xi_\parallel-\xi_\perp)\t\ra\t\ra\right)\cdot\left(\frac{\partial \r\ra}{\partial t}-\v\rba\right),\slabel{1a}\\
\f\rb=-\left( \xi_\perp\I + (\xi_\parallel-\xi_\perp)\t\rb\t\rb\right)\cdot\left(\frac{\partial \r\rb}{\partial t}-\v\rab\right),
\end{subeqnarray}
where all fields are implicitly functions of $s$ and $t$ and where 
 $\xi_\perp$ and  $\xi_\parallel$ are the drag coefficients  for motion in the direction perpendicular  and parallel to its local tangent \cite{gray55,cox70,lighthill75}. 
 We  compute below the hydrodynamic force density acting on filament \#1, the other one being deduced by symmetry.  
 In Eq.~\ref{1a}, the term  $\v ^{(2)\to(1)}$ denotes the flow induced by the motion of filament $2$ near filament $1$: it represents the effect of hydrodynamic interactions and the  goal of this paper is to show how to calculate its value. As filament \#2 undergoes in general both rotational and  translational  motion, we  split  $\v ^{(2)\to(1)}$  into the flows induced by local moments, $\v_M\rba$ {(rotation)}, and those induced by local forces, $\v_F\rba$ {(translation)}. We 
then  write $
\v\rba=\v_M\rba+\v_F\rba$, 
and  calculate the values of each term in the long-wavelength limit, $ h\ll L$.

In order to  simplify the presentation, we focus in detail on the derivation of the  first velocity term, $\v\rba_M$, induced  by the   rotational motion of filament \#2, while the value of $\v_F\rba$  is computed along similar lines (see below). Note that while    $\v\rba_M$ is exactly zero for non-rotating filaments,  e.g.~in the case of the  planar waving flagella of spermatozoa, it will be important in other situations involving rotation, e.g.~the dynamics of bacterial flagellar filaments. 
Since $a \ll h$, the flow may be described by a superposition of flow singularities. If $\m\rb$ denotes the  hydrodynamic  torque density acting on  filament  \#2, the  flow is given as a line of  integral of rotlets (or point torques)  as  \cite{batchelor70}
\begin{equation}\label{vm}
\v_M\rba(s)=\int_0^L\frac{-\m\rb(s')}{8\pi\mu}\times\frac{\R(s, s')}{R(s, s')^3} \dd s',
\end{equation}
where $s$ and $s'$ are the arclengths along filaments   $\#1$ and $\#2$ and $\R(s, s')=\r\ra(s)-\r\rb(s')$ is the relative position vector  with magnitude $R$ (all quantities are  implicit functions of time).  If  filament \#2 rotates  relative to the background fluid with rotation   rate  $\omega\rb(s')$ then it is a classical result that
{
\begin{equation}
\m\rb(s')=-\xi_r\omega\rb(s')\t\rb(s'),
\end{equation}}
where the resistance coefficient in rotation  is $\xi_r=4\pi\mu a^2$. 

We nondimensionalize lengths by $L$, leading to two dimensionless numbers:  the filament aspect ratio, $\epsilon_a=a/L$, and the distance-to-size ratio, $\epsilon_h=h/L$. Times are non-dimensionalised by a  relevant, problem-specific  time scale $T$. 
The integral from Eq.~\ref{vm} becomes then  
in dimensionless form
\begin{align}
\label{vmND}
\bar{\v}_M\rba(\bar{s})=\frac{\epsilon_a^2}{2}\int_0^1 \bar{\omega}(\bar{s}')\rb\t\rb(\bar{s}')\times\frac{\bar{\R}(\bar{s},\bar{s}')}{\bar{R}(\bar{s},\bar{s}')^3}\dd \bar{s}',
\end{align}
and we drop the bars for notation convenience.

\begin{figure}[t]
\centering
\includegraphics[width=0.45\textwidth]{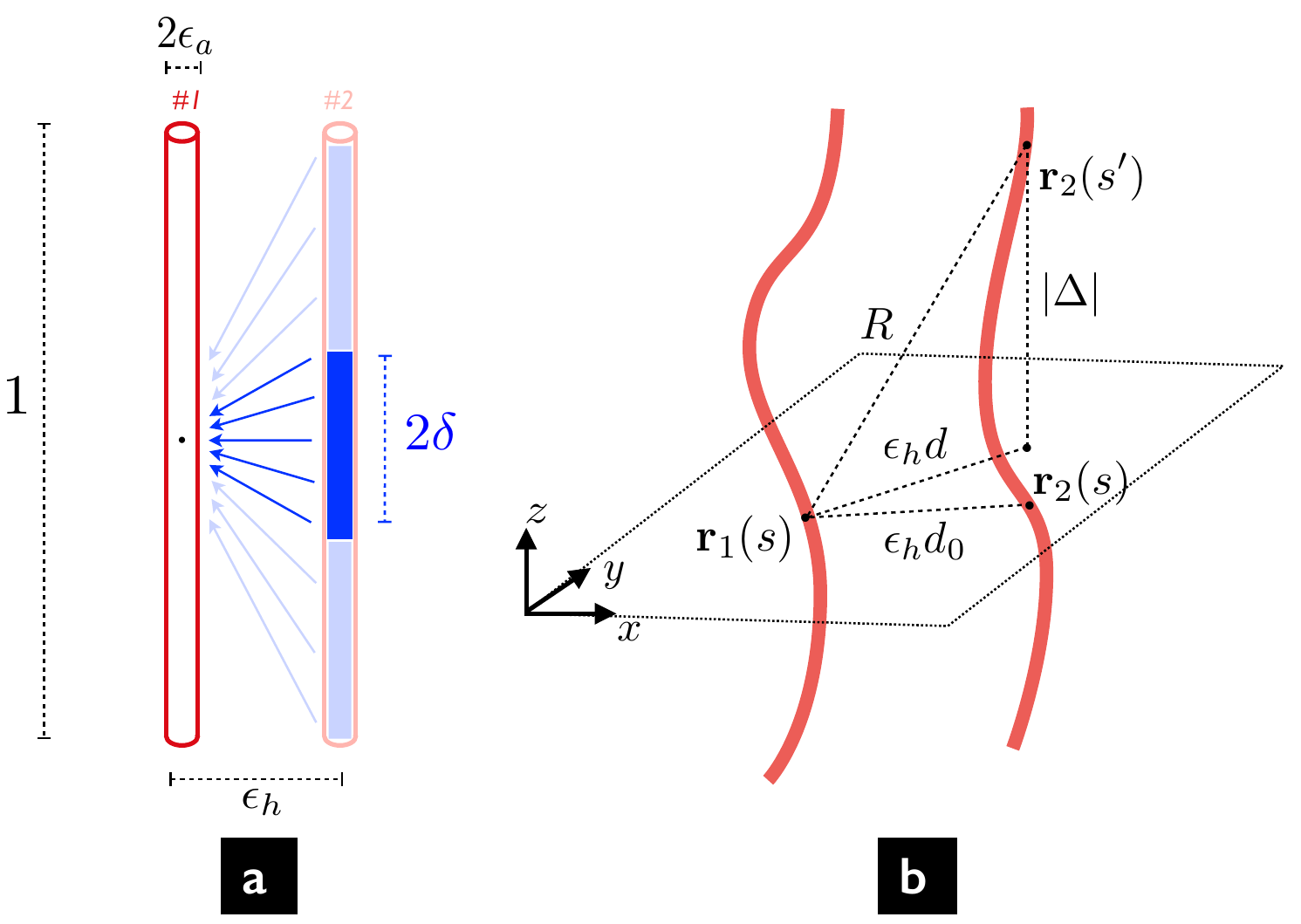}
\caption{(a) Illustration of method to compute hydrodynamic interactions. The integration region is split into a local region of size $2\delta$ and a non-local region where the separation of lengths $\epsilon_h\ll\delta\ll 1$ may be exploited to compute the flow separately. (b) Geometric relations between $d$, $d_0$ and $R$: $R$ is the distance between points on two filaments,  $\epsilon_hd$ the projection in $x-y$ plane and $\epsilon_hd_0$  the local separation distance.}\label{fig2}
\end{figure}

{Since we are in the long-wavelength limit,   it is natural to use cartesian coordinates (Fig.~\ref{fig2}).}  
We denote  by $\z$  the unit vector along the mean direction of the (approximately) parallel filaments and describe the  instantaneous geometry of each filament  as
$\r\ri(t, s^{(i)}) =\left [\epsilon_hx\ri(t, s^{(i)}), \epsilon_hy\ri(t, s^{(i)}), s^{(i)}\right]$ where $s^{(1)}\equiv s$ and $s^{(2)}\equiv  s'$.  Introducing  the notation $\Delta =s-s'$ and the planar vector
$
\d(s;s')=[x\ra(s)-x\rb(s'), y\ra(s)-y\rb(s'), 0] 
$ 
of magnitude  $ d=|\d|$,  then  the relative position vector $\R$ is written by separating the direction along and perpendicular to the filaments as 
$\R=\Delta\z+\epsilon_h\d,$ with magnitude 
$R=\left(\Delta^2+\epsilon_h^2d^2\right)^{1/2}$.

The schematic representation of how the integration is performed is shown in Fig.~\ref{fig2}a with detailed notation in Fig.~\ref{fig2}b. Our method is inspired by a classical calculation due to Lighthill where, in order to describe the flow induced by the motion of a single filament, he separated the flow induced by point singularities into  local and nonlocal terms  using an intermediate  length scale on which the filament was still slender but almost straight \cite{lighthill75}. 
We   introduce an intermediate length scale $\delta$  satisfying $\epsilon_h\ll\delta\ll 1$ and  split the integration into two regions: (1) a  nonlocal region,  $|\Delta|\geq \delta$,  where the distance between two points on the filaments is dominated by  $R\sim |\Delta|$  since $\epsilon_h\ll\delta$ ({resulting velocity denoted $\v^{{\rm NL}}$}); and (2) a local region where $|\Delta|\leq\delta$, and for which in the limit $\delta \ll 1$ we can approximate $R\sim\left(\Delta^2+\epsilon_h^2d_0^2\right)^{1/2}$ where $d_0$ is the local {filament-filament} distance $d_0(s)=d(s;s'=s)$ (resulting velocity  denoted $\v^{{\rm L}}$). The final result, sum of 
$\v^{{\rm NL}}$ and $\v^{{\rm L}}$,  should then be independent of the value of $\delta$.

Changing the variable of integration in Eq.~\ref{vmND}  to $\Delta=s-s'$,  the non-local contribution to the integral is given by
\begin{align}\label{vMNL}
\v_M^{{\rm NL}}=\frac{\epsilon_a^2}{2}\left(\int_{s-1}^{-\delta}+\int_{\delta}^{s}\right)
\left[\omega\rb\t\rb\right]_{(s-\Delta)}\times\frac{\R}{R^3}\dd\Delta. 
\end{align}
Since $|\Delta|\geq \delta$ and    $\epsilon_h\ll\delta$, we have $R(s;s-\Delta) \approx |\Delta|$. Writing $\R=\Delta\z+\epsilon_h\mathbf{d}(s;s-\Delta)$ and  
$\t\rb(s-\Delta)=\z+\epsilon_h\t_\perp(s-\Delta)$ where
$\displaystyle\t_\perp(s-\Delta)=\left(\frac{\partial{x\rb}}{\partial{s'}}, \frac{\partial{y\rb}}{\partial{s'}},0\right)\big\vert_{s'=s-\Delta}$, 
the integrand from Eq.~\ref{vMNL} is   given by 
\begin{equation}\label{12}
\epsilon_h\omega\rb(s-\Delta)
[ \z\times \mathbf{d}(s;s-\Delta) +\t_\perp(s-\Delta)\times \Delta\z]
\frac{1}{|\Delta|^3}\cdot
\end{equation}
The leading-order term  in Eq.~\ref{12}  diverges as $1/\Delta^3$ in the limit $\delta \to 0$, leading to a  final  asymptotic integral as
\begin{equation}\label{6}
\v_M^{{\rm NL}}=\frac{\epsilon_a^2\epsilon_h}{2}\left(
\int_{-\delta}^{s-1}
\!\!\!+\int_{\delta}^{s}\right)
\omega\rb(s-\Delta)
 \z\times \mathbf{d}(s;s-\Delta)
\frac{\dd\Delta}{\Delta^3}\cdot
\end{equation}
In the limit where $\delta \to 0$, the result in Eq.~\ref{6}  diverges and is dominated by the behavior of the integrand near the boundary, i.e.~$\Delta=0$. Calling
$\d_0$  the local direction between the filaments perpendicular to  their long axis, i.e.~$\d_0=\mathbf{d}(s;s)$ (Fig.~\ref{fig2}b), we  obtain  in the limit $\delta \to 0$ 
\begin{align}\label{vMNLresult}
\v_{M}^{{\rm NL}}=\frac{\epsilon_a^2\epsilon_h}{2\delta^2}\omega\rb(s)\z\times\d_{0},
\end{align}
at leading order.

{Next we consider the local integration  where  we have
\begin{align}\label{vML}
\v_M^{{\rm L}}=\frac{\epsilon_a^2}{2}\int_{-\delta}^{\delta}
\omega\rb(s-\Delta)\t\rb(s-\Delta)\times \frac{\R}{R^3}d\Delta.
\end{align}
In the local region, we can Taylor-expand  $\omega\rb$ and $\d$ around $\Delta=0$ 
(i.e.~around  $s'=s$) as
\begin{equation}
\left(
\begin{array}{c}
\displaystyle  \omega(s-\Delta)\rb      \\
\displaystyle \d(s-\Delta)     
\end{array}
\right)
=
\left(
\begin{array}{c}
\displaystyle  \omega(s)\rb      \\
\displaystyle \d(s)       
\end{array}
\right)
+\Delta
\left(
\begin{array}{c}
\omega\rb_{0\Delta}        \\
\d_{0\Delta}       
\end{array}
\right)+O(\Delta^2),
\end{equation}
where, under the long-wavelength approximation, the  derivatives 
$\omega\rb_{0\Delta}$ and 
$\d_{0\Delta} $ 
are  of order one (i.e.~the geometry and the rotation of the filaments vary on the  length scale $L$).  In that case,  each term in the integrand can be expanded
and we get at leading order  that only the local values of the rotation rate, $\omega^{(2)}(s)$, and the  force, $\f\rb(s)$, enter the problem, with a local flow given by
\begin{align}
\label{vMLresult1}
\v_{M}^{{\rm L}}=\frac{\epsilon_a^2\epsilon_h}{2}\omega\rb(s)\int_{-\delta}^{\delta}\frac{\z\times\d_0}{(\Delta^2+\epsilon_h^2d_0^2)^{\frac{3}{2}}}\dd\Delta,
\end{align}
which may  be evaluated analytically  with an asymptotic expression given by
\begin{align}\label{vMLresult2}
&\v_{M}^{{\rm L}}=\frac{\epsilon_a^2\epsilon_h}{2}\omega\rb(s)\,\z\times\d_0
\left(\frac{2}{\epsilon_h^2d_0^2}-\frac{1}{\delta^2}\right).
\end{align}

Adding up Eq.~\ref{vMNLresult} and \ref{vMLresult2},  we obtain the final flow induced by filament \#2, which is independent of the value of $\delta$, given at leading-order by
\begin{equation}\label{vMresult}
\v_M\rba=\frac{\epsilon_a^2}{\epsilon_hd_0^2}\omega\rb(s)\z\times\d_{0}.
\end{equation}

A similar approach may  be used to evaluate the second velocity term, $\v\rba_F$, induced by the  forcing of filament \#2 on the fluid.  In that case, the flow is given by  a line integral of  stokeslet  singularities  (point forces) as
\begin{equation}\label{vf}
\v_F\rba(s)=\int_0^L\frac{-\f^{(2)}(s')}{8\pi\mu}\cdot\left(\frac{\I}{R}+\frac{\R\R}{R^3}\right)\dd s',
\end{equation}
where $\I$ is the identity tensor and  $\f\rb$ the    force density  acting on filament \#2. 
 One notable difference between Eq.~\ref{vm} and Eq.~\ref{vf} is that the integrand in  Eq.~\ref{vm}  is known explicitly (filament rotation), whereas that in  Eq.~\ref{vf} has in it the   quantity we are trying to determine, specifically the  unknown force density, $\f^{(2)}$. We can however proceed as above as long as $\f^{(2)}$ varies on the  length scale $L$, and similarly for the other filament, so that   the resulting velocities in Eq.~\ref{fh} will lead to a linear system to invert to determine both $\f^{(1)}$ and $\f^{(2)}$.  After  nondimensionalising  force densities by $8\pi\mu L/T$, the nonlocal contribution of the integral in Eq.~\ref{vf} is written as
 \begin{equation}
\v_F^{{\rm NL}}=-\left(\int_{s-1}^{-\delta}+\int_{\delta}^{s}\right)\left(\frac{\I}{R}+\frac{\R\R}{R^3}\right)\cdot\f^{(2)}(s-\Delta)\dd\Delta,
\end{equation}
whose evaluation at leading-order value is given by the  logarithmic term
\begin{equation}
\label{vFNLresult}
\v_F^{{\rm NL}}=2 (\ln \delta) (\I+\z \z)\cdot \f\rb(s).
\end{equation}
Similarly, the local portion of the integral, written as 
\begin{equation}
\v_F^{{\rm L}}=-\int_{-\delta}^{\delta}\left(\frac{\I}{R}+\frac{\R\R}{R^3}\right)\cdot\f^{(2)}\dd\Delta,
\end{equation}
 can be  Taylor-expanded and exactly integrated to lead to the local logarithmic dependence
 \begin{equation}
\label{VFL_final}
\v_{F}^{{\rm L}} =2\ln \left(\frac{\epsilon_h d_0}{\delta}\right)(\I + \z\z)\cdot \f\rb(s).
\end{equation}
Adding Eq.~\ref{vFNLresult} and \ref{VFL_final}  we  obtain the final force term as
\begin{equation}
\label{vFresult}
\v_F\rba=
2\ln\left({\epsilon_h d_0}\right)
(\I + \z\z)\cdot \f\rb(s),
\end{equation}
for the velocity induced by the unknown force density.

Returning to dimensional quantities Eqs.~\ref{vMresult}-\ref{vFresult}   can be written as 
\begin{align}\label{vMresult_dimensional}
&\v_M\rba=\left(\frac{a}{h(s)}\right)^2\omega\rb(s)\z\times\h{(s)},\\
&\label{vFresult_dimensional}
\v_F\rba=
\frac{1}{4\pi \mu}\ln\left(\frac{h(s)}{L}\right)
(\I + \z\z)\cdot \f\rb(s),
\end{align}
where  $\h(s)$ is  the dimensional local  vector  between the filaments, i.e.~$\h (s)= \r_1(s)-\r_2(s)$, and $h(s)$ its norm. 

The results in Eqs.~\ref{vMresult_dimensional}-\ref{vFresult_dimensional}, together with Eq.~\ref{fh}  are the main new results of this paper. They provide   a linear, local relationship between the force density on each filament ($\f\ri$) and the kinematics of their motion ($\omega\rj$ and $\partial \r^{(k)}/\partial t$).   As a remark, we note that one is not allowed to formally take the limit $h\to0$ or $h\to\infty$ in Eqs.~\ref{vMresult_dimensional}-\ref{vFresult_dimensional}, as both violate the    limit $a \ll h \ll L$ in which these formulae were derived.  

For planar motion ($\omega\rj=0$ for $j=1,2$), the algebra simplifies further.  In Eq.~\ref{fh}, since $h\ll L$, the tangent vectors are $\t  = \z$ at leading order in $h/L$ 
and since $\xi_\perp\approx 2 \xi_\parallel$ \cite{lighthill75} we have for each filament
\begin{equation}
\xi_\perp\I + (\xi_\parallel-\xi_\perp)\t\t\approx 
\xi_\perp\left( \I-\frac{1}{2}\z\z \right)\equiv \bf J,
\end{equation}
\def\J{{\bf J}}
 so that on each filament $i$ we have the dynamic balance
\begin{equation}\label{balance}
\f\ri(s,t) - \J\cdot \v^{(j)\to(i)}=-\J\cdot \frac{\partial \r\ri}{\partial t},
\end{equation}
with $j\neq i$. 
{Given the tensorial operator appearing in  Eq.~\ref{vFresult_dimensional}, we have to evaluate 
\begin{equation}
\left( \I-\frac{1}{2}\z\z \right)\cdot 
(\I + \z\z)=\I,
\end{equation}
and we further note that 
$
{\xi_\perp}/{4\pi\mu }\approx 
{1}/{\ln (1/\epsilon_a)}
$ \cite{lighthill75}. 
As a result,}   Eq.~\ref{balance}  simplifies for each filament  to
\begin{equation}\label{final}
\f^{(i)}(s,t) + \frac{\ln({h(s,t)}/{L})}{\ln (a/L)}\f^{(j)}(s,t)=-\J\cdot \frac{\partial \r^{(i)}}{\partial t},
\end{equation}
with $j\neq i$.  Defining $\lambda(s,t)\equiv  {\ln({h(s,t)}/{L})}/{\ln (a/L)}$ and $\Lambda (s,t)\equiv 1-\lambda^2(s,t)$ (note that $\Lambda >0$ since $a < h$), 
  this linear system can be inverted by hand and we obtain the analytical formula for the force density $\f^{(i)}(s,t)$ acting on filament $i$ as 
\begin{equation}\label{final2}
\f^{(i)}(s,t)=-\frac{1}{\Lambda(s,t)}\J\cdot \left(\frac{\partial \r^{(i)}}{\partial t}-\lambda(s,t) \frac{\partial \r^{(j)}}{\partial t}\right).
\end{equation}

We now illustrate predictions of our theory  on a simple model of two rigid filaments undergoing  planar motion, and compare with  numerical slender-body simulations. Consider two straight  coplanar filaments of radius $a$, length $L$  with centerlines located at  $[0,\epsilon y_1(z,t),z]$ and $[0,h_0+\epsilon y_2(z,t),z]$. Assume for simplicity small amplitude  motion $\epsilon\ll 1$ and let us use our results to calculate the force density in the $y$ direction, $f^{(i)}=\f^{(i)} \cdot  \e_y$, in powers of the amplitude ($f^{(i)}=\epsilon f_1^{(i)} + \epsilon^2 f_2^{(i)}+...$) in the limit $a\ll h_0\ll L$.    
Writing  $h=h_0 + \epsilon h_1$,  a Taylor expansion gives
{\begin{equation}
\ln \left({h}/{L}\right)=\ln \left({h_0}/{L}\right) + \epsilon{h_1}/{h_0} + O(\epsilon^2),
\end{equation}}
which we use to evaluate  Eq.~\ref{final2} at order $\epsilon$, leading to
\begin{align}\label{invert_order1}
f_1^{(i)} =\frac{\xi_\perp}{1-[\ln(h_0/L)/\ln(a/L)]^2}
\left(\frac{\ln(h_0/L)}{\ln(a/L)}
\frac{\partial y^{(j)}}{\partial t}
-\frac{\partial y^{(i)}}{\partial t}\right).
\end{align}
At order $\epsilon^2$, Eq.~\ref{final} becomes
\begin{equation}\label{order2}
f_2^{(i)} + \frac{\ln({h_0}/{L})}{\ln (a/L)}f_2^{(j)}=-\frac{h_1}{h_0\ln (a/L)}f_1^{(j)},
\end{equation}
Assuming that both $y_1$ and $y_2$ are periodic in time on the same period, then a time-average of Eq.~\ref{order2} using Eq.~\ref{invert_order1} leads to identical  mean force densities along both filaments as $
\big\langle f_2^{(1)} \big\rangle= 
\big\langle f_2^{(2)} \big\rangle=f_2(s)$, 
where   
\begin{equation}\label{fs}
f_2(s)=\frac{\xi_\perp \ln(a/L)}{2h_0[\ln(h_0/L)+\ln(a/L)]^2}
\bigg\langle
y_-\frac{\partial y_+}{\partial t}
\bigg \rangle,
\end{equation}
with  $y_+\equiv y^{(1)}+y^{(2)}$
and $y_-\equiv  y^{(2)}-y^{(1)}$. 

For illustration purposes, let us assume that the first filament undergoes sinusoidal rigid-body motion  of the form
$y_1(t)={\cal R}(\sum_n \ell_n \exp in \omega t )$ while the second filament has the same motion with a phase difference $\phi$,  
i.e.~$y_2(t)=y_1(t+\phi)$. Our theory, Eq.~\ref{fs},  predicts that the two-rod system will pump the fluid by exerting a net force on it, $F$,  of magnitude
\begin{equation}
F_2=\frac{4\pi\mu \omega L }{2h_0[\ln(h_0/L)+\ln(a/L)]^2} \sum_n n |\ell_n|^2\sin (n\phi)\cdot\label{39}\end{equation}
Clearly Eq.~\ref{39} predicts zero  net force for in-phase ($\phi=0$) and out-of-phase ($\phi=\pi$) motion and thus an optimal phase difference between the two filaments exits.  

We test in Fig.~\ref{fig3} this theoretical prediction  against a numerical implementation of nonlocal slender-body 
 appropriate for interactions \cite{johnson80,tornberg06} in the case $n=1$.  
 We numerically solve for the force distribution along each filament using a Galerkin method based on Legendre polynomials.   The net force on each filament  is then computed at 15 equidistant points within a period, and the mean force calculated. While the theoretical approach (Eq.~\ref{39})  was derived only asymptotically in the limit where $a/h\to0$ and  $h/L\to0$  we see that even when these  parameters  are not asymptotically small (here $a/h=0.25$ and $h/L=0.1$), the theoretical prediction (solid line) is able to capture the computational results (dashed  line and symbols) with good approximation. In contrast, far-field predictions are off by more than two orders of magnitude. 

\begin{figure}[t]
\centering
\includegraphics[width=0.45\textwidth]{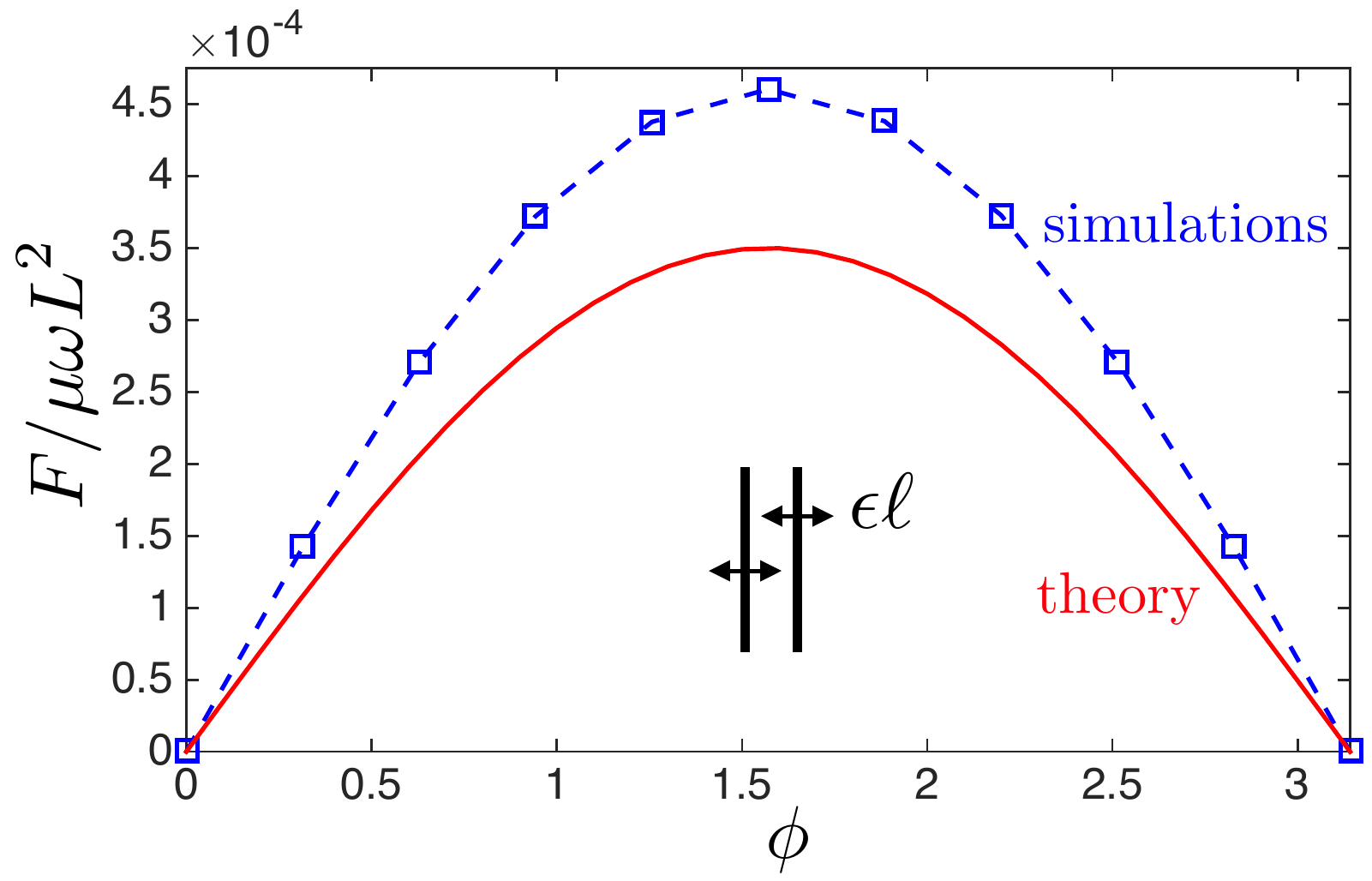}
\caption{
Net force induced on the fluid by a two-rod pump, $F/\mu \omega  L^2$, as a function of the phase difference, $\phi$, between the   rods. Dashed line and symbols: slender-body theory simulations; Solid line: theory (Eq.~\ref{39}). The dimensionless distance between the rods is   $h/L=0.1$, their aspect ratio    $a/L=0.025$ (so that $a/h=0.25$) and the  motion amplitude is $\epsilon \ell = h/10$.}
\label{fig3}
\end{figure}

In summary, we have used an asymptotic method to compute the hydrodynamic interactions between nearby filaments undergoing arbitrary rotation and translation. The key ingredient allowing the calculation to be carried out  is to exploit the separation of length scales $a\ll h \ll L$ which enables a representation of the flow as a superposition of fundamental singularities whose strengths vary only on  long wavelengths compared to the separation between the filaments.  
While the work above was derived only in the case of  filaments with main directions parallel to each other,  future work with be required to  generalise the results  to the case of  non-parallel filaments; we speculate that the ``local'' aspect of the final equation is likely to involve the point on each filament which is nearest to the other.

Like  any other asymptotic derivation, a crucial question in our work is that of the magnitude of the error (i.e.~the order of the next-order terms)}.  
To fix ideas, consider first {a single filament} undergoing planar deformation with a centerline described by $[x,y(x,t)]$. The classical formula for the leading-order force density, $\f$, on  the filament is $\f=-(\xi_\perp\partial y / \partial t)\e_y$, with (i)  logarithmic corrections in the aspect ratio of the filament from next-order terms beyond resistive-force theory, 
i.e.~relative error $O(1/\ln (L/a))$ \cite{cox70}  and (ii) algebraic corrections in the typical slope of the filament, i.e.~relative error $O(h/L)$ due to the difference between the true instantaneous geometry of the filament and its mean direction \cite{lauga09}. {The same relative errors apply to our current work}.  Additional errors arise in our work near the ends of the filaments. Specifically, in order for the non-local integrations to be carried out near the ends of the filaments, the arclength $s$ needs to satisfy  $h \ll \min(s, L-s)$, with logarithmically (resp.~algebraically) small relative errors in $ h / \min(s, L-s)$  from filament translation (resp.~rotation). {Physically, this logarithmic accuracy of local hydrodynamics is the equivalent to that of resistive-force theory but extended to  multiple filaments.}   {The portion of the filament with an admissible  arclength $s$ satisfying $\min(s,L-s)\gg h$ is of  size $L-2s_0$, with $s_0 \gg h$. Since we are in the limit $h\ll L$,  the geometric mean $s_0=\sqrt{hL}$ satisfies the intermediate limit $L\gg s_0 \gg h$. As a consequence,  our results are  able to  provide  the value of the hydrodynamic force density on the majority of the filaments, namely at least a portion of size $L-2\sqrt{hL}$. 
}

We finally point out that while the addition of   higher-order flow singularities than rotlet and stokeslets along each filament would improve the analysis,  the resulting additional terms would decay spatially  algebraically and faster than the terms in Eqs.~\ref{vMresult_dimensional}-\ref{vFresult_dimensional}, which  provide thus the leading-order contribution in the limit $a \ll h \ll L$.

The framework developed in this paper will  allow to address theoretically a number of problems in the biomechanics of filaments where  nonlocal hydrodynamic interactions may be integrated out analytically for example in cytoskeletal mechanics, hydrodynamic interactions and cellular propulsion, beyond the classical,  complementary, far-field approach.  {For example, two particular problems  in the realm of biological synchronisation \cite{goldstein2016elastohydrodynamic}
 could be tackled: the requirements  for attraction and synchronisation between the rotating helical flagellar filaments of 
bacteria~\cite{Powers04,Stark05} and the generation of metachronal waves in cilia 
arrays~\cite{brennen77,guirao07,Niedermayer08}.}

{Our results  should also be applicable to a broad range of problems in physical sciences where slender bodies interact through a viscous fluid, such as  liquid crystals. As an example, a set of recent measurements showed   strong  interactions between living organisms and a liquid crystal~\cite{zhou14}, a situation which could be addressed using our framework.}

\section*{Acknowledgments}
We thank Ray Goldstein and Adriana Pesci for useful discussions, in particular on their earlier work on a similar calculation. This work was funded in part by the European Union through a CIG grant and a ERC Consolidator grant to EL and by the Cambridge Trust. 
\bibliography{HI_refs}
\end{document}